\documentclass[prd,reprint,preprintnumbers,twocolumn,notitlepage]{revtex4-2}

\usepackage{multirow}
\usepackage{siunitx}

\usepackage{tikz}
\usetikzlibrary{decorations.markings, arrows, positioning, calc}

\usepackage{graphicx,times}
\usepackage{latexsym}
\usepackage{mathtools}

\usepackage{amsmath,amssymb,amsbsy,amsfonts}
\usepackage{array}
\usepackage{bm}
\usepackage{graphics}
\usepackage{mathrsfs}
\usepackage{xcolor}
\usepackage{cancel}
\usepackage[normalem]{ulem}

\usepackage[unicode, pdfprintscaling=None, colorlinks]{hyperref}
\hypersetup{
    colorlinks,
    allcolors=blue!50!black,
}
\usepackage{inconsolata}

\usepackage[capitalise]{cleveref}

\newcommand{\RR}{{\mathbb{R}}}

\renewcommand\>{\rangle}
\newcommand\<{\langle}

\newcommand\Order{O}

\usepackage{relsize}

\usepackage{microtype}

\newcommand{\cc}[2]{c_{{#1}/{#2}}}

\usepackage{ctable}

\newcommand\Config{\mathcal{C}}

\newcommand\LT{L_{T}}

\newcommand\Zbig{Z_M}
\newcommand\Zworm{Z_{Q, w}}
\newcommand\Zmeas{\tilde Z_{Q+1}}
\newcommand\Zdef{Z_{Q}}

\newcommand\SiteSink{(\LT/2, \vec 0)}
\newcommand\SiteSource{(0, \vec 0)}
\newcommand\ZmeasXT{\Zmeas^{(t, \vec x)}}

\newcommand\Qmin{Q_{\text{min}}}
\newcommand\Qmax{Q_{\text{max}}}
\usepackage{relsize}

\usepackage[acronyms, smallcaps, nohypertypes={acronym}, nopostdot, style=super, nonumberlist, toc]{glossaries}
\glsenableentrycount
\makeglossaries  
\setacronymstyle{long-sc-short}

\newcommand\ac[1]{\gls{#1}}

\newcommand\acp[1]{\glspl{#1}}
\newcommand\Acp[1]{\Glspl{#1}}

\newacronym{WF}{wf}{Wilson-Fisher}
\newacronym{AF}{af}{asymptotically free}
\newacronym{RG}{rg}{renormalization group}
\newacronym[longplural={conformal field theories}]{CFT}{cft}{conformal field theory}
\newacronym[longplural={lattice field theories}]{LFT}{lft}{lattice field theory}
\newacronym[longplural={effective field theories}]{EFT}{eft}{effective field theory}
\newacronym[longplural={quantum field theories}]{QFT}{qft}{quantum field theory}
\newacronym{LEC}{lec}{low-energy constant}
\newacronym{QCD}{qcd}{quantum chromodynamics}
\newacronym{MC}{mc}{Monte Carlo}
\newacronym{IR}{ir}{infrared}
\newacronym{UV}{uv}{ultraviolet}
\newacronym{SNR}{snr}{signal-to-noise ratio}
\newacronym{NLSM}{nlsm}{nonlinear sigma model}
\newacronym{CSA}{csa}{Cartan subalgebra}
\newacronym{SSB}{ssb}{spontaneous symmetry breaking}
\usepackage{xfrac}

\begin{document}

\title{Large-charge conformal dimensions at the $O(N)$ Wilson-Fisher fixed point}
\author{Hersh Singh}
\preprint{IQuS@UW-21-022,INT-PUB-22-008}
\email{hershsg@uw.edu}
\affiliation{InQubator for Quantum Simulation (IQuS), Department of Physics, University of Washington, Seattle, Washington 98195-1550, USA}

\begin{abstract}
  Recent work using a large-charge expansion for the $O(N)$ Wilson-Fisher conformal field theory has shown that the anomalous dimensions of large-charge operators can be expressed in terms of a few low-energy constants (LECs) of a large-charge effective field theory (EFT). By performing lattice Monte Carlo computations at the $O(N)$ Wilson-Fisher fixed point, we compute the anomalous dimensions of large-charge operators up to $N=8$ and charge $Q=10$, and extract the leading and subleading LECs of the $O(N)$ large-charge EFT. To alleviate the signal-to-noise ratio problem present in the large-charge sector of conventional lattice formulations of the $O(N)$ theory, we employ a recently developed qubit  formulation of the $O(N)$ nonlinear sigma models with a worm algorithm. This enables us to test the validity of the large-charge expansion and the recent large-$N$ predictions for the coefficients of the large-charge EFT.
\end{abstract}

\maketitle

\section{Introduction}
\label{sec:intro}

\Acp{CFT} are ubiquitous in physics. In condensed matter physics, they describe the critical behavior of materials at second-order phase transitions \cite{zinn-justin_quantum_2021, cardy_scaling_1996}, and in particle physics, they arise naturally as \ac{RG} fixed points of relativistic \acp{QFT} \cite{komargodski_constraints_2012, polchinski_scale_1988, wilson_renormalization_1983}. Understanding \acp{CFT} is therefore pivotal to understanding a wide range of physical phenomenon.

Despite their importance, \acp{CFT} remain a challenge to study
This is because, in general, \acp{CFT} are strongly-coupled and it can be difficult to find a small parameter to do a perturbative expansion.
Nonetheless, much progress has been made by exploiting ``hidden'' small parameters such as in the large-$N$ expansion \cite{moshe_quantum_2003}, the $\epsilon$-expansion \cite{wilson_critical_1972},
or with non-perturbative techniques such as conformal bootstrap \cite{poland_conformal_2019, simmons-duffin_conformal_2016}.
In some cases, lattice field theory methods offer a completely non-perturbative method to compute properties of a \ac{CFT} numerically \cite{pelissetto_critical_2002}.
All these techniques have their limitations and provide a window into a potentially different regimes of the \ac{CFT}. 

In the last few years, there has been progress using a new small parameter: the inverse of a global charge $Q$ \cite{hellerman_cft_2015,gaume_selected_2020}.
For theories with a global symmetry, we can study the theory in sectors of fixed global charge $Q$ and obtain an expansion in inverse charge $Q^{-1}$.
The work of Ref.~\cite{hellerman_cft_2015} 
showed that, in certain theories such as the $O(2)$ \ac{WF} \ac{CFT}, restricting to sectors of fixed global charge  causes a spontaneous breakdown of the global symmetry, giving rise to Goldstone modes \cite{nielsen_how_1976}.
We can then write down a low-energy \ac{EFT} for the Goldstone modes, where the higher-order operators are suppressed in powers of the inverse charge $Q^{-1}$.
The \ac{EFT} description allows us to make several predictions about the behavior the theory in sectors of large global charge.
This was also extended to non-Abelian symmetries such as $O(2n)$ in many recent studies
\cite{alvarez-gaume_compensating_2017, moser_convexity_2022, jack_anomalous_2021, antipin_more_2021, dondi_resurgence_2021, badel_epsilon_2019, antipin_charging_2020, badel_feynman_2020}.

While \acp{EFT} encode information about symmetries and relevant degrees of freedom, they also involve \acp{LEC} which are \emph{a priori} unknown. 
This prevents a quantitative computation of many observables of interest.
Ideally, we would like to be able to compute the \acp{LEC} directly from the underlying theory.
The only known general technique which enables such quantitative computations in field theories is numerical lattice field theory, typically using \ac{MC} methods.
Since lattice \ac{MC} computations can work directly with the underlying theory, they can also provide independent check of the \ac{EFT} methods.

This two-pronged approach is also commonly used, for example, in studies of \ac{QCD}.
Since the low-energy sector of \ac{QCD} is strongly coupled, we can write down \acp{EFT} such as chiral perturbation theory or pionless \ac{EFT} \cite{bedaque_effective_2002},
describing the physics of low-energy degrees of freedoms, with unknown \acp{LEC}.
One then computes the \acp{LEC} from lattice \ac{QCD} computations, which allows both a non-trivial test of the \ac{EFT} methods, as well as a way to turn the low-energy \acp{EFT} into a powerful quantitative tool.

It is conceivable that such an combined lattice and \ac{EFT} approach can also be used with large-charge \acp{EFT} to study strongly-coupled \acp{CFT}, especially if a sign-problem free formulation of the \ac{CFT} on the lattice can be found.
This approach was taken in the works of Refs.~\cite{banerjee_conformal_2018, banerjee_conformal_2019, banerjee_subleading_2021} for the $O(2)$ and $O(4)$ \ac{WF} \acp{CFT}.
The authors tested the predictions of large-charge \ac{EFT} using lattice \ac{MC} methods.
In particular, a prediction of the large-charge \ac{EFT} for these models is that the conformal dimensions of leading charge-$Q$ operators admit an expansion in $Q^{-1}$,
\begin{align}
  D(Q) = \cc32 Q^{3/2} + \cc12 Q^{1/2} + c_0 + \Order(Q^{-1/2}),
  \label{eq:DQ}
\end{align}
where $\cc32, \cc12, \dotsc $ are the unknown \acp{LEC}.
In Ref.~\cite{banerjee_conformal_2018}, the authors used a lattice formulation of the $O(2)$ \ac{NLSM} to directly compute the conformal dimensions of the charged operators, using which they computed the \acp{LEC} by fitting to \cref{eq:DQ}.
This idea was then applied to the $O(4)$ \ac{WF} \ac{CFT} in Ref.~\cite{banerjee_conformal_2019}.
Surprisingly, the authors observed that, in both these cases, the validity of the large-charge expansion seems to extend all the way down to very small charges $Q \sim 1$.

Given the success of the large-charge expansion for the $O(2)$ and $O(4)$ theories, one might wonder if this feature of the large-charge expansion persists for the $O(N)$ models with larger $N$ as well.
Recently, the authors of Ref.~\cite{alvarez-gaume_large_2019}, studied the $O(N)$ model in a combined large-charge and large-$N$ expansion.
They showed that the expansion given in \cref{eq:DQ} can be obtained for $O(N)$ models as well.
Further, they derived expressions for the large-charge \ac{EFT} coefficients $\cc32, \cc12$ in the large-$N$ limit. 
However, a numerical test of the validity of the large-$N$ results was only available for small $N=2,4$.

In this work, we fill this gap by extending the approach of earlier works \cite{banerjee_conformal_2019, banerjee_conformal_2018} to study large-charge sectors of the $O(N)$ \ac{WF} fixed point for larger $N$ using lattice \ac{MC} techniques.
Using efficient worm algorithms, we compute the conformal dimensions $D(Q)$ of leading charge-$Q$ operators of the $O(N)$ \ac{WF} fixed point, up to $Q=10$ and $N=8$.
Assuming the expansion given in \cref{eq:DQ}, as predicted by Ref.~\cite{alvarez-gaume_large_2019}, we extract the leading coefficients $\cc32$ and $\cc12$.
This lets us test the validity of the large-charge expansion for larger $N$ and the predictions of the large-$N$ expansion. Our final results are summarized in \cref{fig:large-NQ-results} and \cref{tab:results}.

One of the interesting features of the analysis of Ref.~\cite{banerjee_conformal_2019} was the use of a ``qubit'' $O(4)$ model to avoid a sign problem with the traditional $O(N)$ lattice vector model in sectors of large charges.
In Refs.~\cite{singh_qubit_2019, singh_qubit_2019a}, we generalized their $O(4)$ model to $O(N)$ qubit models for arbitrary $N$, and provided evidence to show that these qubit $O(N)$ models have a second-order critical point in the $O(N)$ \ac{WF} universality class for $N=2,3,4,6,8$ in three spacetime dimensions.
Therefore, we can use the qubit model tuned to this critical point for studying the \ac{WF} \ac{CFT}.
From the perspective of lattice \ac{MC} computations, these models are appealing since they admit a worldline representation as a model of $N/2$ (for $N$ even) colored, oriented loops.
In such a worldline formulation, we can develop a worm algorithm which can sample configurations very efficiently and compute the correlation functions, even in the large-charge sectors.
These models have also recently become interesting due to their potential for studying \acp{QFT} on quantum computers \cite{singh_qubit_2019a, buser_quantum_2020, bhattacharya_qubit_2021, singh_qubit_2019, zhou_spacetime_2021, liu_qubit_2022}.

This paper is organized as follows.
In \cref{sec:wilson-fisher}, we review aspects of the large-charge expansion for the $O(N)$ \ac{WF} \ac{CFT} relevant to our work.
In particular, we emphasize the results for the conformal dimensions of the leading charged operators.
In \cref{sec:ON-qubit-model}, we review the $O(N)$ qubit model developed in Ref.~\cite{singh_qubit_2019a}, which we shall use to compute the conformal dimensions of large-charge operators, and describe a Monte Carlo approach based on worm algorithm to efficiently compute the conformal dimensions.
We discuss the results of lattice calculations and compare with analytic results from literature in \cref{sec:results}.
We summarize this work and present our conclusions in \cref{sec:conclusions}.

\section{Large-Charge Expansion for the $O(N)$ Wilson-Fisher Fixed Point}
\label{sec:wilson-fisher}

\subsection{Wilson-Fisher CFT}
We can define the $O(N)$ \ac{WF} \ac{CFT} as the continuum limit of a certain lattice model at a second-order critical point.
For example, we may consider the the partition function $ Z= \int \mathcal{D \phi} \ e^{-S_L}$ with the Euclidean lattice action
\begin{align}
  S_{L} = - \beta \sum_{\< x y \>} \vec \phi_x \cdot \vec \phi_y
  \label{eq:trad-model}
\end{align}
defined on a $D$-dimensional Euclidean lattice with periodic boundary conditions.
Here, $\vec \phi_{x} \in \RR^{N}$ is an $N$-component bosonic real scalar field constrained such that $|\vec \phi_x| = 1$, the sum runs over all bonds $\<xy\>$ between nearest-neighbor lattice sites $x$ and $y$, and $\beta$ is a coupling that we can tune to a critical point.
The field $\vec \phi_x$ transforms in the fundamental representation of $O(N)$, so that this lattice action has a manifest global $O(N)$ symmetry.

In $D \geq 3$ spacetime dimensions, this lattice model is known to have a second-order critical point at $\beta=\beta_c$ (finite).
This critical point describes the spontaneous breakdown of the global $O(N)$ symmetry down to $O(N-1)$.
As we make $\beta$ large and positive, the fields align and the system transitions from a disordered, $O(N)$ symmetric phase into an ordered phase with broken $O(N)$ symmetry.
In $D=4$ spacetime dimensions, the corresponding \ac{RG} fixed point is the Gaussian fixed point,  which just gives the free field theory in the continuum limit with logarithmic corrections.
However, in $D=3$, this is a nontrivial fixed point called the \ac{WF} fixed point \cite{wilson_critical_1972}. The continuum limit at this critical point is an interacting conformally-invariant field theory, called the \ac{WF} \ac{CFT}.

\subsection{Large-charge expansion}

Since the goal of this work is to obtain quantitative results on the $O(N)$ Wilson-Fisher \ac{CFT} and to connect with the large-charge expansion, let us briefly review the results of the large-charge expansion for the $O(N)$ model. 
We highlight the results which are relevant to our work and refer the reader to the original literature for details \cite{alvarez-gaume_compensating_2017, hellerman_cft_2015, alvarez-gaume_large_2019}. 

The question that concerns us here is: what is the conformal dimension $D(Q)$ of the leading primary operator in the $O(N)$ \ac{WF} {CFT} with a given charge $Q$?
For the Abelian $O(2)$ group, the charge $q$ of a operator $\mathcal{O}_q$ determines the transformation under an $O(2)$ rotation in the field space, parametrized by an angle $\theta$ as $\mathcal{O}_q \to e^{i q \theta} \mathcal{O}_q$.
In the limit of large $Q$, the analysis of Refs.~\cite{alvarez-gaume_large_2019, alvarez-gaume_compensating_2017, hellerman_cft_2015} shows that, the  conformal dimensions $D(Q)$ admit an expansion in inverse powers of $Q$,
\begin{align}
  D(Q) = \cc32 Q^{3/2} + \cc12 Q^{1/2} + c_0 + \Order(Q^{-1/2}) ,
  \label{eq:D-Q-ON-large-charge}
\end{align}
where $\cc32, \cc12, c_0, \ldots$ are some unknown coefficients.

This notion of a charge can be generalized to $O(2n)$ by fixing  $n$ orthogonal planes and considering independent $O(2)$ rotations in each plane.
This form the maximal torus subgroup of $O(2n)$.
Let the $n$ independent $O(2)$ rotations be parameterized by angles $\vec \theta = (\theta_1, \dotsc, \theta_n) $,
where $\theta_i$ parametrizes the $O(2)$ rotation in the $(2i-1, 2i )$ plane.
A operator $\mathcal{O}_{\vec q}$ with well-defined $O(2n)$ charges transforms as $\mathcal{O}_{\vec q} \to e^{i \vec \theta \cdot \vec q} \mathcal{O}_{\vec q}$ under the action of generators in the Cartan subgalebra, where we define the vector $\vec q = (q_1, \dotsc, q_n  )$ as the \emph{charge}  of the operator $\mathcal{O}_{\vec q}$.
In the large-charge expansion, the relevant quantity is 
$Q = \sum_{i}|q_i|$.
With this definition of $Q$, the authors of Ref.~\cite{alvarez-gaume_large_2019} found that \cref{eq:D-Q-ON-large-charge} holds for the $O(2n)$ theory as well.

The result in \cref{eq:D-Q-ON-large-charge} is obtained from an \ac{EFT} analysis.
In sectors of fixed non-zero global charge, the authors of Ref. \cite{alvarez-gaume_large_2019} note that the global $O(2n)$ symmetry
is explicitly broken to a $U(n)$ symmetry, which further undergoes a spontaneous breakdown to $U(n-1)$:
\begin{equation}
  O(2n) \xrightarrow{\text{explicit}} U(n) \xrightarrow{\textsc{ssb}} U(n-1).
\end{equation}
This is a slight generalization of the familiar scenario of \ac{SSB} to sectors of fixed global charge, and indeed we find Nambu-Goldstone modes
\cite{nielsen_how_1976}.
There are $(n-1)$ Goldstone bosons with a non-relativistic dispersion relation, and a single relativistic Goldstone boson.
At low-energies, one can write down an \ac{EFT} description of the relativistic Goldstone boson, where the higher-order operators are suppressed by powers of inverse charge $Q^{-1}$.

Being an \ac{EFT} result, the \acp{LEC} are unknown and depend on $N$.
This is true for all the \acp{LEC} except $c_0$, which is predicted to have a universal (independent of $N$) value of $c_0 \approx -0.0937$.
Recently, the combined large-$N$ and large-charge analysis of Ref.~\cite{alvarez-gaume_large_2019} predicted the coefficients $\cc32, \cc12$ as functions of $N$,
\begin{align}
  \cc32 = (2/3)  N^{-1/2},\quad
  \cc12 = (1/6)  N^{1/2},
  \label{eq:large-N}
\end{align}
which should be valid in the regime $Q \ll N \ll Q^2$ for large $Q,N$.

\Cref{eq:D-Q-ON-large-charge,eq:large-N} are the focus of this work.
We would like to develop a numerical \ac{MC} method to accurately compute the conformal dimensions $D(Q)$, and test of validity of \cref{eq:D-Q-ON-large-charge,eq:large-N} for the $O(N)$ theories over a wide range of $N$ and charge $Q$.  We do this in the following sections.

\section{$O(N)$ Wilson-Fisher CFT from a qubit lattice model}
\label{sec:ON-qubit-model}

\subsection{Qubit $O(N)$ models}

\begin{figure*}[htbp]
  \centering
  \includegraphics[width=0.9\linewidth]{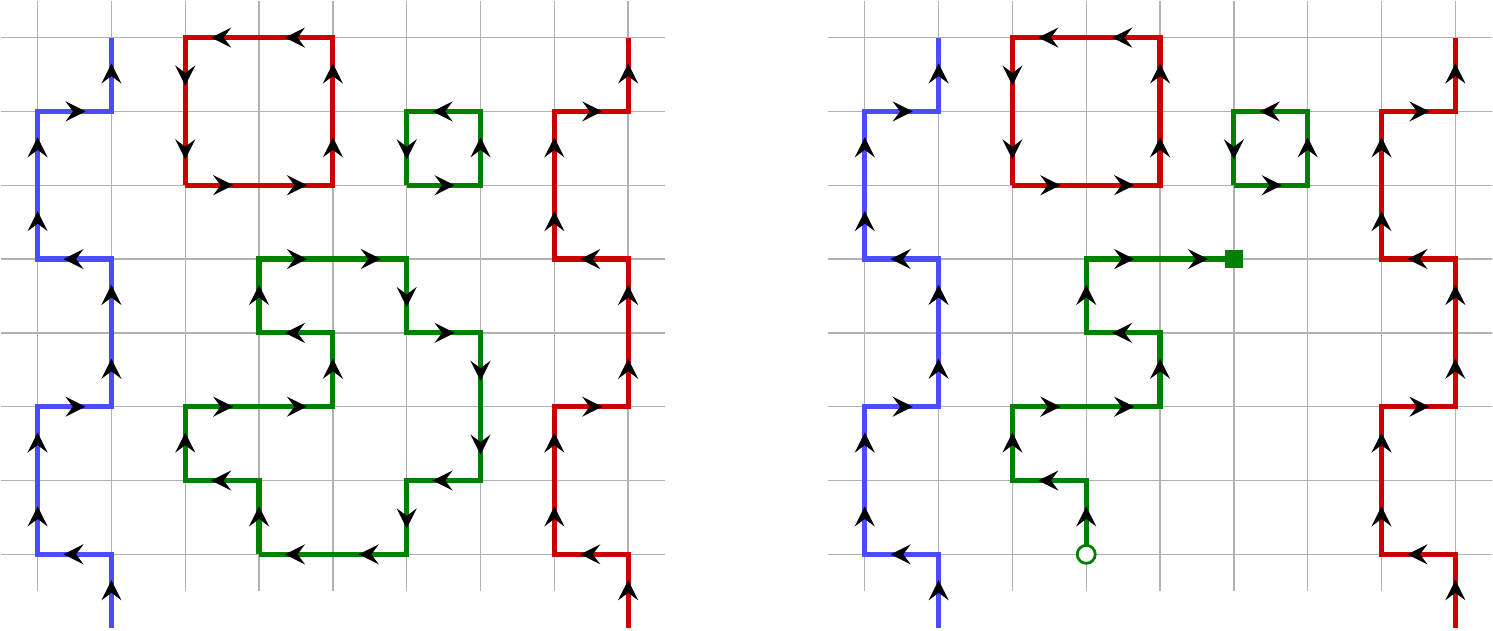}
  \caption{%
    Typical worldline configurations contributing to partition function for the $O(6)$ qubit model in the default sector (left) and in the worm sector (right).
    The worldline representation for the $O(6)$ qubit model has 3 types of oriented worldlines, each representing a different $O(6)$ charge, which we show with different colors.
  }
  \label{fig:worldline-worm}
\end{figure*}

The $O(N)$ \ac{NLSM}  may be obtained as the continuum limit of the lattice action in \cref{eq:trad-model}.
While there are efficient cluster algorithms for the traditional $O(N)$ model \cite{wolff_collective_1989} for performing computations close to zero charge density, we are interested in numerical computations in sectors of large global $O(N)$ charges.
In such cases, traditional \ac{MC} approaches using the lattice action in \cref{eq:trad-model} suffer from a sign problem.

In order to develop an efficient \ac{MC} approach, we will instead use an $O(N)$ model with a finite-dimensional local Hilbert space, which we call a ``qubit'' $O(N)$ model.
This model was constructed in a recent work \cite{singh_qubit_2019a} as an $O(N)$ generalization of the $N=2,3,4$ models studied in Refs.~\cite{cecile_modeling_2008, banerjee_finite_2010, banerjee_conformal_2019, singh_qubit_2019}.
The key point is that, in $(2+1)$ dimensions, this model has a second-order critical point in the $O(N)$ $\ac{WF}$ universality class.
Therefore, we use this model at criticality to extract the conformal dimensions by measuring appropriate correlation functions.

Here, we quickly review the model and refer the reader to Ref.~\cite{singh_qubit_2019a} for more details.
The local (single-site) Hilbert space for the model is an $(N+1)$-dimensional space constructed as the direct sum of a singlet representation (which we think of as the empty ``Fock vacuum'' state) and an $N$-dimensional fundamental representation (which we think of as $N$ ``particle'' states) of the $O(N)$ group.
The finite-temperature partition function of this model, $Z = e^{-\beta H}$, can be written as a worldline formulation in a manifestly spacetime symmetric way:
\begin{align}
  Z = \sum_{\Config} W[\Config],
  \label{eq:Z-worldline}
\end{align}
where the sum is over worldline configurations $\Config$ on a periodic Euclidean spacetime lattice, and $W[\Config]$ is the weight of the configuration $\Config$.

For the qubit $O(2n)$ model defined on a periodic spacetime lattice, we find that the configurations are composed of non-intersecting, closed, oriented worldlines of $n$ ``colors.''
An oriented worldline with color $i$ (where $i=1,\dotsc, n$) represents a state with $O(2n)$ charge $\vec q = \pm \hat e_i$ (where $\vec e_i$ is an $n$-vector with one at the $i$the position and $0$ elsewhere). That is,  the color $i$ worldlines have charge $\pm 1$ under $O(2)$ rotations in the $(2i-1, 2i)$ plane, and charge zero for rotations in all other planes.
There are $n$ such worldlines, corresponding to the $n$ types of particles/anti-particle pairs transforming under the fundamental representation of $O(2n)$.
For illustration, a typical configuration contributing to the partition function for the $O(6)$ model is shown in the left panel of \cref{fig:worldline-worm}, where we see oriented worldlines of 3 colors.

This spacetime symmetric model has only one parameter $U$, which is the weight of each bond between two sites.
At $U=0$ there are no particles and the partition function is dominated by the trivial configuration with only singlets. At very large and positive $U$, the dominant contributions come from the $O(2n)$ particles, resulting in an $O(2n)$ symmetric ground state.
However, in $3d$, we expect the $O(2n)$ symmetry of the ground state to be broken down to $O(2n-1)$.
Therefore, as we tune $U$ from $0$ to $\infty$, 
this model goes to through a second-order phase transition at some $U=U_c$ which corresponds to the \ac{SSB} scenario $O(2n)\to O(2n-1)$. This critical point was located precisely in Ref.~\cite{singh_qubit_2019a}, and confirmed to lie in the $O(2n)$ \ac{WF} universality class by computing the critical exponents. Therefore, by tuning the model to  this critical point $U=U_c$, we can study the physics of the \ac{WF} fixed point.

\begin{figure}[t]
  \centering
  \includegraphics[width=0.99\linewidth]{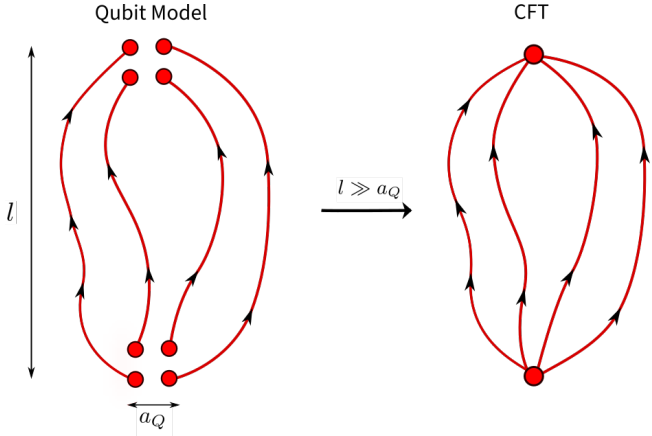}
  \caption{Computing the large-charge correlators of the \ac{CFT} from the qubit model. In the qubit model, we place $Q$ sources and sinks of a given color, each of unit charge, over a small region of linear size $a_Q$. In the limit of large separation $l \gg a_Q $, this computes the \ac{CFT} correlation function of the leading charge-$Q$ operators. }
  \label{fig:large-charge-scaling}
\end{figure}
\subsection{Conformal dimensions from the qubit $O(N)$ model}
\label{sec:conformal-dimensions}

In this section, we describe a numerical \ac{MC} method to compute the conformal dimensions of the charged operators.
The method was introduced in Ref.~\cite{banerjee_conformal_2019} for the $O(4)$ model, but easily generalizes to the $O(N)$ case. 

We are interested in the computing two-point correlators of the form
\begin{align}
  \< \phi_Q(x)\ \phi_{-Q}(y) \> \sim |x-y|^{-2 D(Q)}
\end{align}
where $\phi_Q$ is the leading primary operator with a charge $Q$, and $D(Q)$ is its conformal dimension.
However, the qubit $O(N)$ model does not have single-site operators with charge $Q>1$.
At any given site, the allowed charged states lie in the fundamental representation with $Q=1$.

We can work around this issue by appealing to scaling arguments.
The charge-$Q$ correlators of the \ac{CFT} can be accessed in the qubit model by spreading $Q$ unit charges, say $\vec q_1 = (1,0,\dotsc,0)$, over a small region to construct a smeared charge-$Q$ excitation over that region. 
To be concrete, let $\hat \phi_{L}(x)$ be any single-site lattice operator with charge $\vec q_{1}$ at the spacetime site $x$.  Let $x_1,\dotsc,x_Q$ be $Q$ spacetime points all within a radius $\sim a_Q$ of the point $x$.
We define a product of lattice operators
\begin{align}
  \Phi_{Q}^{L}(x) = \phi^{L}(x_1) \cdots \phi^{L}(x_Q),
\end{align}
which has a well-defined $O(2n)$ charge $\vec q = (Q, 0, \dotsc, 0)$.
At criticality, the long-distance behavior of the two-point function of the lattice operators $\< \Phi_{Q}^{L}(x) \Phi_{-Q}^{L}(y) \>$ gets a dominant contribution from the two-point function of the leading primary charge-$Q$ operators,
\begin{align}
  G_Q(x, y) = \< \Phi_{Q}^L(x) \Phi_{-Q}^L(y) \> &\sim |x - y|^{-2D(Q)},
                                           \label{eq:charge-correlator}
\end{align}
in the limit $|x-y| \gg a_Q$.
This is shown schematically in \cref{fig:large-charge-scaling}.
Since we are interested only in the universal exponent of the leading power-law term, the exact form of the smearing does not matter.

\subsection{Sketch of the Monte Carlo method}

We would like to now develop an \ac{MC} algorithm to measure correlators of the form in  \cref{eq:charge-correlator}, in the presence of charged sources and sinks.
We begin by defining a partition function for worldline configurations in a given charge sector,
\begin{align}
  Z_Q(L) = \sum_{\Config_Q} W[ \Config_Q ],
  \label{eq:Z-Q}
\end{align}
where we have explicitly specified the dependence on the system size $L$,
and the sum is over worldline configurations $\Config_Q$ having $Q$ sources and sinks placed in a specific manner as follows.
Our model is manifestly spacetime symmetric, but we arbitrarily label one of the directions as time for convenience, and denote a spacetime point by $(t, \vec x)$, where $\vec x$ is a spatial site on the time slice $t$.
Let $L_T$ be the extent of the box in the time direction.
We use sources and sinks of the same charge, say, $\vec q_1 = (1,0,\dotsc,0)$.
On the $t=0$ time slice, we place $Q$ sources in a small region close to $\vec x = 0$, and on the $t=\LT/2$ time slice, we place $Q$ sinks around the site $\vec x=0$.

\begin{figure}[tbp]
  \centering
  \includegraphics[width=0.9\linewidth]{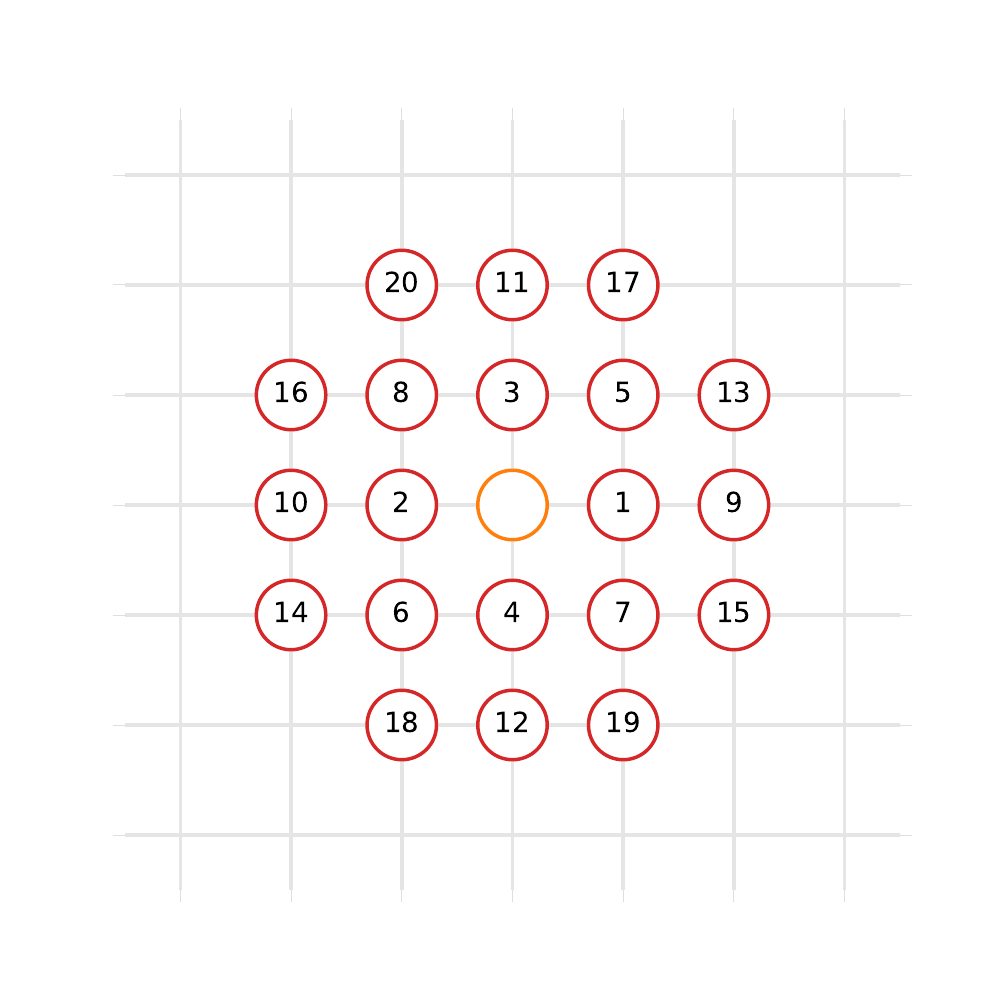}
  \caption{Placement of the sources and sinks for $Z_{Q}$ at the two-dimensional spatial slices $t=0$ and $t=\LT/2$.
    Each circle denotes a source (at $t=0$) or a sink (at $t=\LT/2$) of charge $\vec q_1 = (1, 0,\dotsc, 0)$.  The numbers denote the order in which we place these charges as we increase the total charge $Q$. For example, the charge placement shown in this figure corresponds to $Q=20$.
    The central site (corresponding to $\vec x = 0$) is left empty for $Z_Q$, but is used in when adding an additional charge for the measurement update while sampling $\tilde Z_{Q+1}^{(\vec x, t)}$.
  }
  \label{fig:source-config}
\end{figure}

\begin{figure}[htbp]
  \centering
  \includegraphics[width=0.8\linewidth]{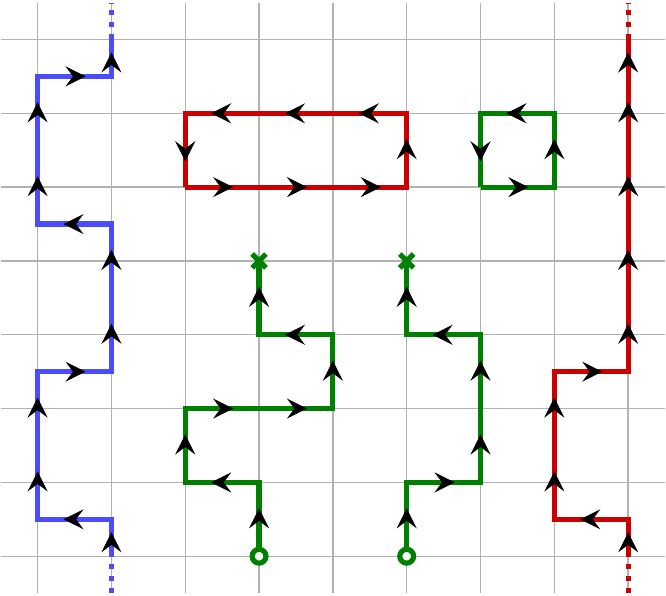}
  \caption{A typical worldline configuration for the $O(6)$ model with three colors, and two sources and sinks for green worldlines.
    The open circles indicate sources, and the crosses indicate sinks.  The site between the two source sites is left empty as it is used in the measurement update.}
  \label{fig:worldline-source}
\end{figure}

The exact placement of the charges does not matter in the scaling regime.
However, the spread of these charges does introduce a short distance scale $a_Q \sim \sqrt{Q}$ in the system.
Therefore, we choose the placement to have the smallest spread $a_Q$ possible.
We show our choice in \cref{fig:source-config}, where each circle shows a site with a unit charge $\vec q_1$, and the number showing the order in which these charges are placed as we increase the total charge of the system $Q$.
As long as we have large separation between the sources and sinks, $l = L_T/2 \gg a_Q$, we expect the universal power-law decay of the \ac{CFT} to emerge.
(Note that for configurations contributing to $Z_Q$, we leave the site $\vec x=0$ empty on the both on the source and sink time slices. This special site will be used in an additional step below.)
A typical configuration contributing to $Z_Q$ is shown in \cref{fig:worldline-source}, in the $Q=2$ sector.

Let us now define another partition function $\ZmeasXT(L)$ in the charge $(Q+1)$ sector in the following way. 
The first $Q$ sources and sinks are placed in exactly the same way as for $Z_Q$ specified above.
However, now there is an additional source at the spacetime site $(0, \vec 0)$ and a sink at the arbitrary site $(t, \vec x)$.
Finally, we define an extended partition function,
\begin{align}
  \Zbig = Z_{Q} + \sum_{(t, \vec x)} \tilde Z_{Q+1}^{(t, \vec x)}.
  \label{eq:Zbig}
\end{align}
We will describe, in \cref{sec:worm-algo}, a  worm algorithm which can efficiently sample configurations in $\Zbig$ by switching between the $Z_Q$ and $\Zmeas^{(t, \vec x)}$ sectors.
This lets us compute the ratio of partition functions  
\begin{align}
  R_{Q}(L) = \frac{\Zmeas^{(L/2, \vec 0)}(L)}{Z_{Q}(L)}.
  \label{eq:RQ}
\end{align}
In the scaling limit $L \gg a_Q$, the ratio $R_{Q}$ in fact computes the ratio of two-point correlators of leading charge-$(Q+1)$ and charge-$Q$ operators,
\begin{align}
  R_{Q}(L) \sim \frac{G_{Q+1}(L/2)}{G_Q(L/2)} = c L^{-2\Delta (Q+1)}, \label{eq:ratio-Q}
\end{align}
where we have defined $\Delta(Q)$ as the difference of conformal dimensions
\begin{align}
  \Delta(Q) = D(Q) - D(Q-1).
  \label{eq:Delta-Q}
\end{align}
We can therefore measure $D(Q)$ for any $Q$ by doing a series of computations for different charge sectors to measure $\Delta(Q)$ and computing,
\begin{align}
  D(Q) = \sum_{q=1}^{Q}\Delta(q).
\end{align}
where we set $D(0)=0$.

\subsection{Details of the worm algorithm}
\label{sec:worm-algo}

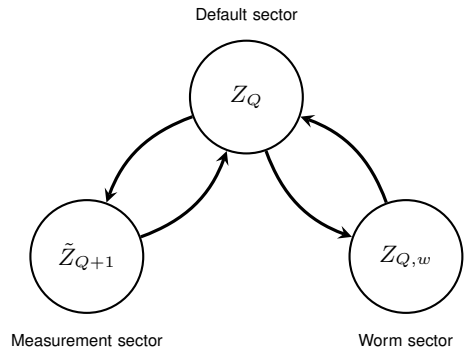
\begin{figure}[htbp]
  \centering
  \begin{tikzpicture}[font=\sffamily,
    sector/.style = {circle, thick, draw, minimum size=1.5cm},
    arrow/.style = {-stealth, bend right=25, very thick},
    desc/.style = {text width=3cm, align=center, font=\relsize{-2}\sffamily}]
    \node[sector] (A) at (0,0) {$Z_Q$};
    \node[sector] (B) at (-45:3) {$\Zworm$};
    \node[sector] (C) at (225:3) {$\Zmeas$};
    \path (A) edge[arrow] (B);
    \path (B) edge[arrow] (A);
    \path (A) edge[arrow] (C);
    \path (C) edge[arrow] (A);
    \node[above=1.5mm of A, desc] {Default sector};
    \node[below=1.5mm of B, desc] {Worm sector};
    \node[below=1.5mm of C, desc] {Measurement sector};
  \end{tikzpicture}
  \caption[Worm algorithm for conformal dimensions]{
    The full worm algorithm for the extraction of conformal dimensions of charge-$Q$ operators.
    We begin in the default sector $\Zdef$, defined as the sum over worldline configurations with $Q$ sources and sinks, each of charge $\vec q_1$, placed at time slices $t=0$ and $t=\LT/2$ as shown in \cref{fig:source-config}.
    We perform worm updates to sample $\Zworm$, which consists of worldline configurations with an additional source and sink (worm tail and head).
    When the worm update ends, we obtain a new configuration in $\Zdef$.
    We then perform a measurement update by proposing to enter $\Zmeas$.
    This is a special worm update of fixed color ($\vec q_1$) where the worm tail (source) is placed at $\SiteSource$.
    When this update ends, we obtain a new measurement for the observable $R_{Q}$, defined in \cref{eq:RQ}.
  }
  \label{fig:algo-conformal-dimensions}
\end{figure}

We can now construct an efficient worm algorithm to sample the partition function \cite{prokofev_worm_2010, prokofev_worm_2001}.
The full algorithm is composed of 3 sectors, where a different partition function is sampled, shown
schematically in \cref{fig:algo-conformal-dimensions}. 
The first is the ``default'' sector, given by \cref{eq:Z-Q}, which consists of configurations in a given charge sector $Q$.
We show an example of such a configuration in the left panel of \cref{fig:worldline-worm}.
This is where we begin the algorithm. 
The other two sectors are the ``worm'' sector $\Zworm$ and the ``measurement'' sector $\Zmeas$.
Starting in the default sector $Z_{Q}$, we propose to either perform a worm update (sampling $\Zworm$), or to perform a measurement update (sampling $\Zmeas$).
The worm update is used to generate new configurations in $Z_{Q}$, and the measurement update is used to measure the observable $R_Q$ defined in \cref{eq:RQ}.
We now describe each of the updates.

\emph{Worm update: }
From the default sector, we first propose to move into the ``worm'' sector, where the partition function $\Zworm$ is sampled.
The configurations contributing to $\Zworm$ are the same as $\Zdef$, except that they also include a single insertions of a creation and an annihilation operator of the same color at any two spacetime sites.
For an example of such a worm configuration, see the right panel of \cref{fig:worldline-worm}.
The key point is that worm algorithms can sample such configurations very efficiently \cite{prokofev_worm_2010, prokofev_worm_2001, evertz_loop_2003}. 
For the zero-charge sector, the details of the worm algorithm used here were described in Refs.~\cite{singh_qubit_2019, singh_qubit_2019a}.
Since we are working in sectors with fixed charge insertions, we modify the algorithm slightly to account for that.

The worm algorithm works by introducing a creation and annihilation operator at a random spacetime site to a configuration in $Z_Q$.
As the worm head moves, it can either create a bond (if it moves towards an empty site), delete a bond (if it moves back on itself), or move a bond (if it moves into an existing worldline). 
The probability of each move is chosen such that the reverse move satisfies detailed balance.
To account for the presence of charged operator insertions (sources and sinks), whenever a proposed move takes the worm head to one of the operator insertions, we simply reject the move.
This ensures that only configurations with a fixed source/sink pattern are sampled.
Once the worm head touches the worm tail, the worm update ends and we are back in the default sector.

\emph{Measurement update:} From default sector $Z_Q$, we then propose to move into the ``measurement sector,'' where partition function $\Zbig$, defined in \cref{eq:Zbig}, is sampled.
This is a worm update very similar to the one defined above, except that we always begin by inserting the worm head/tail at the site $\SiteSource$, instead of a random spacetime site. This site does not contain any sources, by construction of $Z_Q$.
As the worm head moves, it samples configurations contributing to $\Zmeas = \sum_{(t, \vec x)} \Zmeas^{(t, \vec x)}$.
Whenever the worm head touches the spacetime site $\SiteSink$, we count that as a contribution towards $\Zmeas^{\SiteSink}$.
In a given measurement update, let the number of configurations generated in $\Zmeas^{\SiteSink}$ be denoted by $N_{Q+1}$.
The average of $N_{Q+1}$ over a large number of measurement updates computes precisely the observable $R_Q$ defined in \cref{eq:RQ},
\begin{align}
  \< N_{Q+1} \> \approx R_Q = c|\LT|^{-2 \Delta(Q+1)}.
\end{align}
Therefore this observable gives direct access to $\Delta(Q+1)$.  We can repeat this computation over a range of $Q=0, 1, \dotsc, \Qmax$ and compute the conformal dimensions $D(Q)$.

As we increase the lattice size $L$, this algorithm can exhibit a signal-to-noise ratio problem in the measurement sector.
This is because for large lattices, it becomes increasingly unlikely for the measurement worm head to touch the special site $(\vec 0, \LT/2)$ before closing.
To improve the statistics, following Ref.~\cite{banerjee_conformal_2019}, we also perform a reweighting of the worm configurations in $\Zmeas$. 
If the worm tail is at $\SiteSource$ and the worm head is at $(t, \vec x)$, we multiply the weight of the worm configuration by $t^{p}$, where $p > 0$ is chosen empirically.
This makes it easier for the worm to grow very large and increases the likelihood that it touches $\SiteSink$.
It is easy to satisfy detailed balance with this reweighting and it improves the signal-to-noise ratio for the measurement update.

\section{Results}
\label{sec:results}

In this section, we describe our results for the conformal dimensions of leading charge-$Q$ operators for the $O(N)$ \ac{WF} \ac{CFT} for $N=2,4,6,8$ and over a range of $Q$,  and compare with analytic predictions.

\begin{figure*}[htbp]
  \centering
  \def\PlotWidthFactor{0.45}
  \includegraphics[width=\PlotWidthFactor\textwidth]{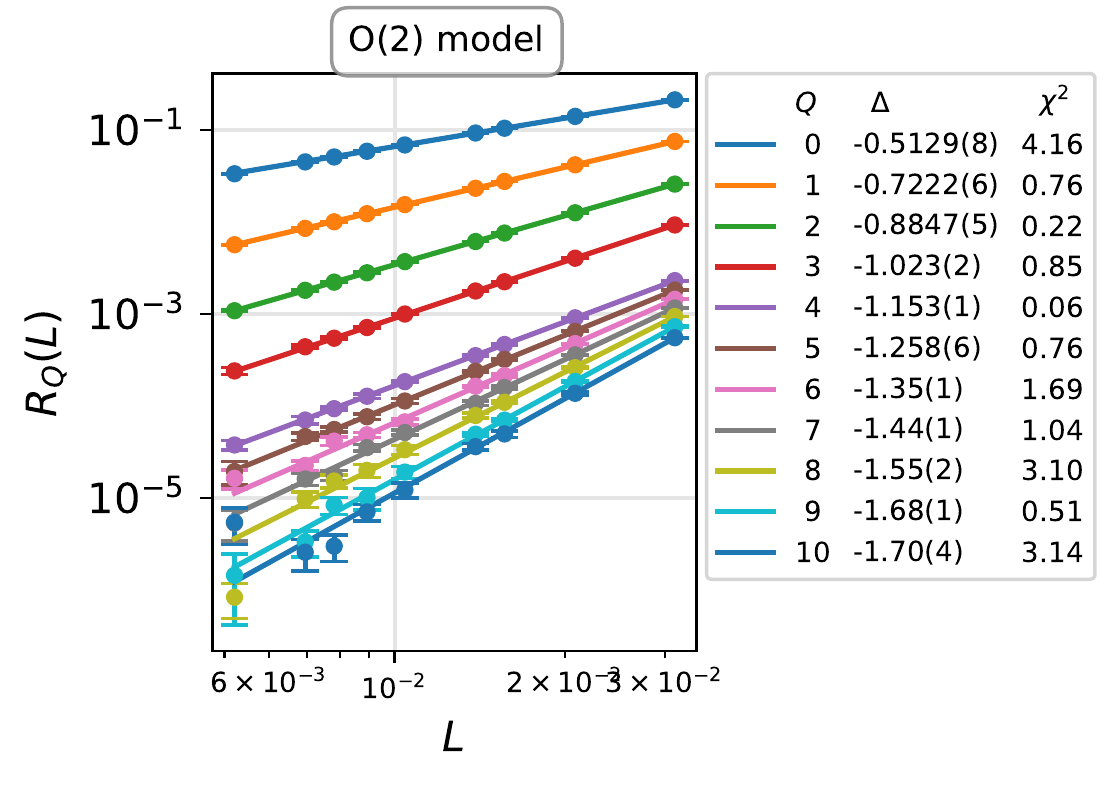}
  \includegraphics[width=\PlotWidthFactor\textwidth]{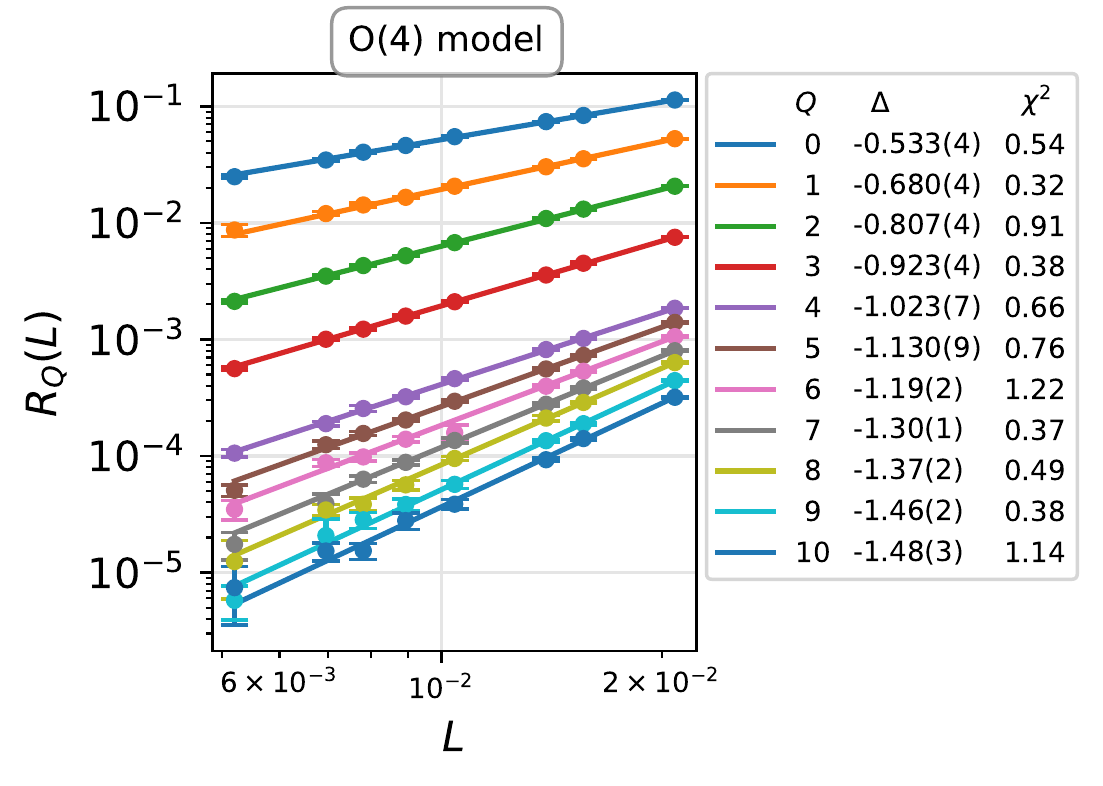}
  \includegraphics[width=\PlotWidthFactor\textwidth]{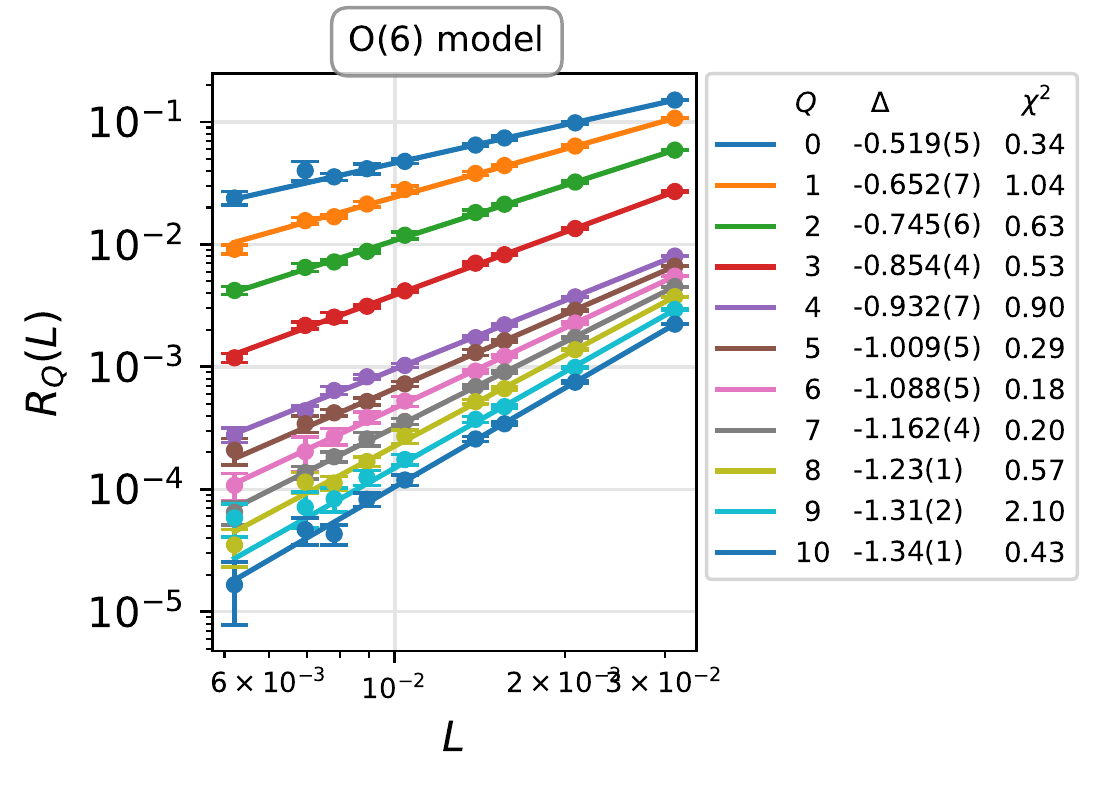}
  \includegraphics[width=\PlotWidthFactor\textwidth]{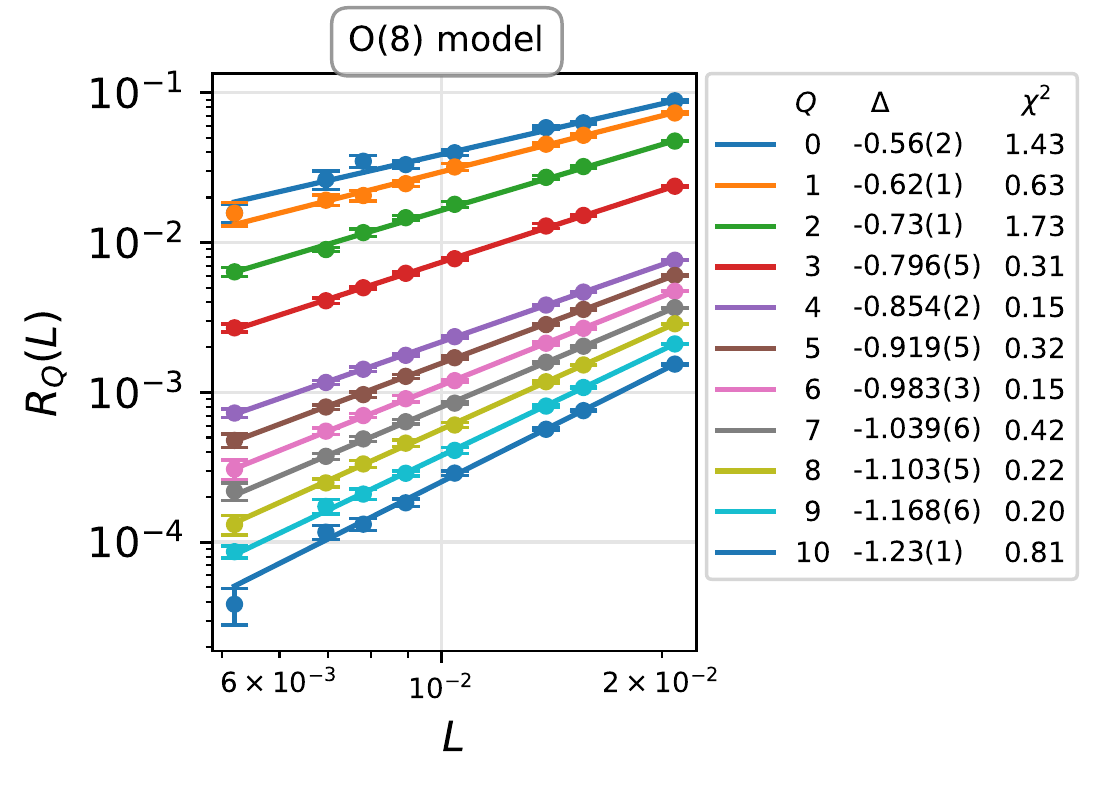}
  \caption{Extraction of conformal dimensions for the charge-$Q$ operators in the $O(N)$ model with $N=2,4,6,8$.
    The solid circles show the lattice \ac{MC} data for the ratio $R_Q(L)$, defined in \cref{eq:RQ}.
    The solid lines are fits to a power-law $R_Q(L) = c L^{-\Delta(Q+1)}$.
  }
  \label{fig:ON-powerlaw-fits}
\end{figure*}

\begin{figure*}[htbp]
  \centering
  \def\PlotWidthFactor{0.48}
  \includegraphics[width=\PlotWidthFactor\textwidth]{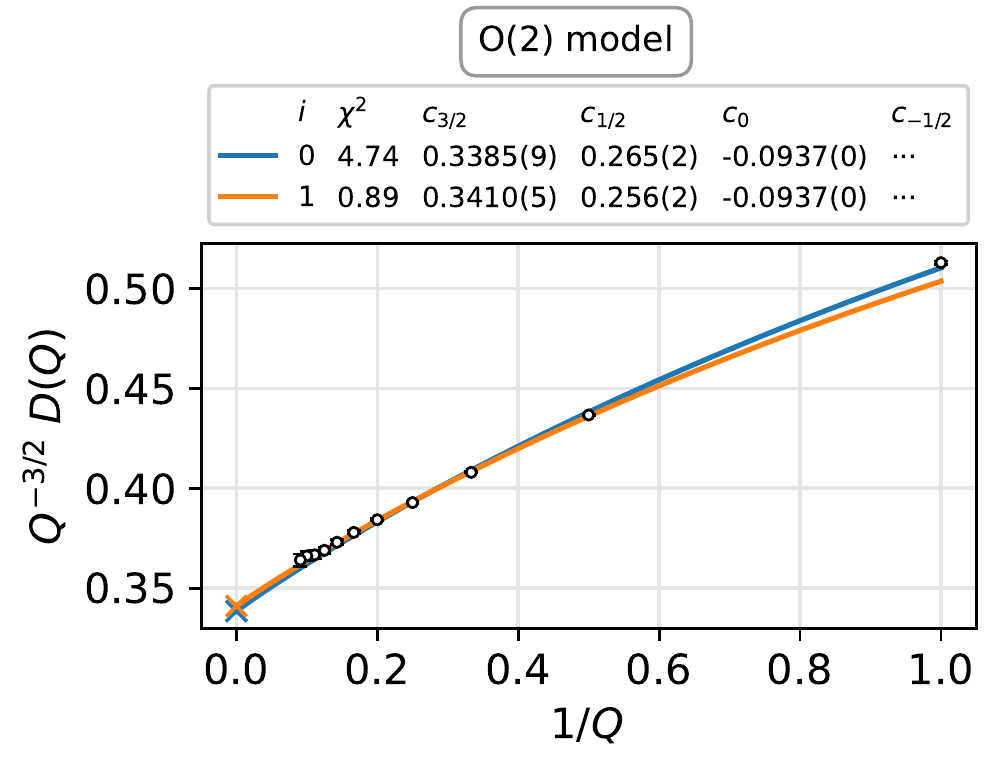}\hfill
  \includegraphics[width=\PlotWidthFactor\textwidth]{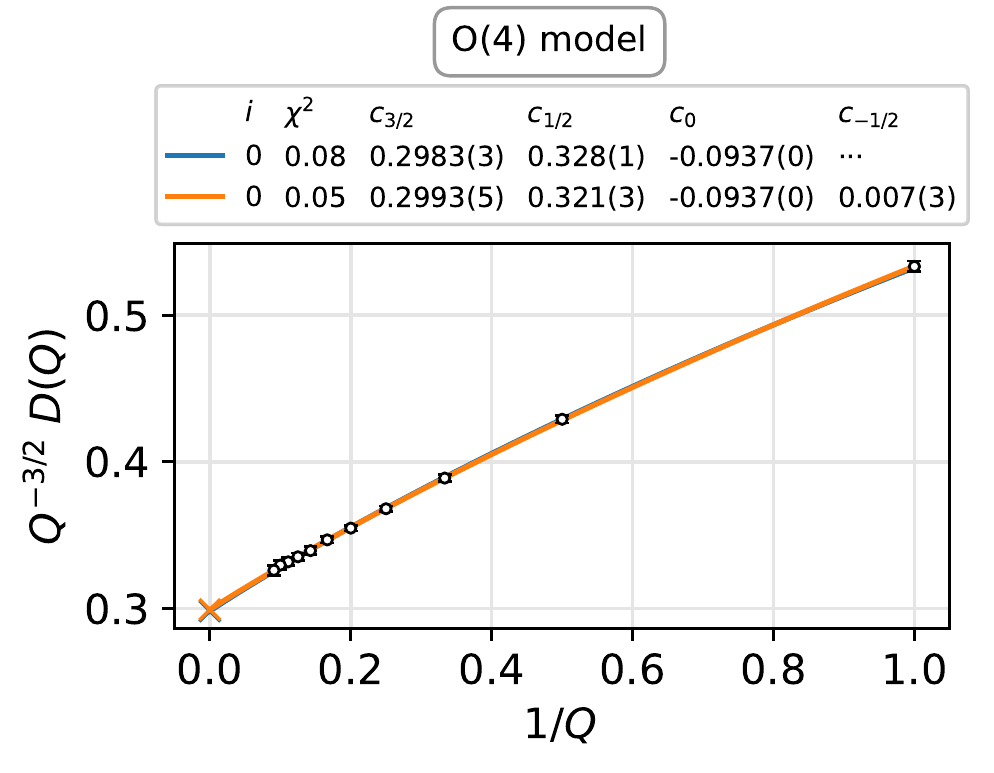}\\
  \includegraphics[width=\PlotWidthFactor\textwidth]{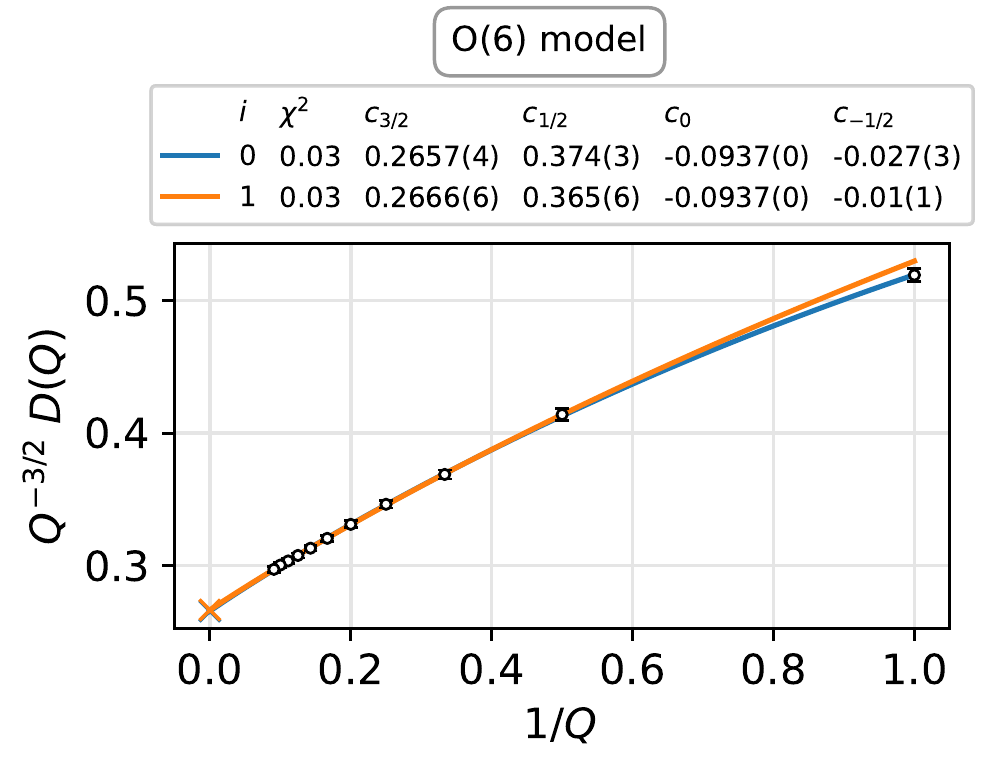}\hfill
  \includegraphics[width=\PlotWidthFactor\textwidth]{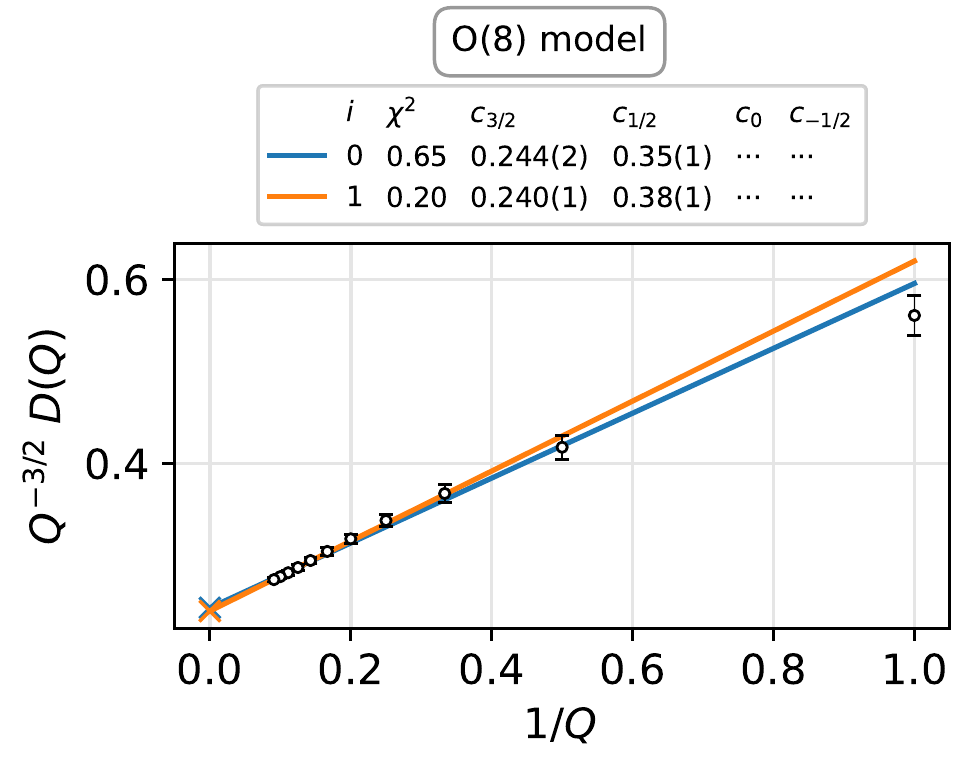}
  \caption[Fits to the large-charge \ac{EFT}]{%
    Extraction of the large-charge \ac{EFT} coefficients from $Q=1,\cdots,10$. 
    We show the results for the $O(N)$ model with $N=2,4,6,8$.
    With $Q^{-3/2}D(Q)$ on the $y$-axis, and $Q^{-1}$ on the $x$-axis, the large-charge \ac{EFT} prediction is that the $x$-intercept is $\cc32$ and the slope at the origin gives $\cc12$.
    The curvature in the data is from the higher order terms ($c_0$ and higher-order) in the large-charge expansion. 
    The small curvature indicates that the large-charge expansion is in fact quite accurate even for very low charge $Q$.
    We perform various fits by excluding the smallest $i$ charges data-points charges to estimate the systematic errors, until $\chi^2/\text{DOF}$ becomes small.
    An ellipsis ($\cdots$) in the columns for coefficient $c_k$ indicates that $c_k$ was not included in the fit, and an error of 0 (for $c_0$) indicates that the value of this parameter was fixed.
  }
  \label{fig:ON-large-charge-fits}
\end{figure*}

\subsection{Extraction of conformal dimensions}

Given a $Q$ and $N$, we perform computations in the charge-$Q$ sector as described in the previous section and compute the observable $R_Q(L)$ defined in \cref{eq:RQ}.
We perform fits of the data to the form
\begin{align}
  R_{Q}(L) = c_Q L^{-2 \Delta(Q+1)}
\end{align}
where $c_Q$ is a non-universal constant and $\Delta(Q)$ is the difference of conformal dimensions defined in \cref{eq:Delta-Q}.
For each $N$, we compute $\Delta(Q)$ for $Q=1,\dotsc,10$.
The conformal dimension $D(Q)$ of the leading charge-$Q$ operator is then extracted by just adding up the differences: $D(Q) = \sum_{q=1}^{Q}\Delta(Q)$.
In \cref{fig:ON-powerlaw-fits}, we show plots of $R_Q(L)$ as a function of $L$ for the $O(N)$ models with $N=2,4,6,8$. Although we find that statistical errors increase slightly with larger $Q$, the data is still precise enough to allow for a clean extraction of the exponent $\Delta(Q)$.

\newcommand\Lmin{L_{\text{min}}}
\newcommand\Lmax{L_{\text{max}}}

\subsection{Extraction of \acp{LEC} of the large-charge \ac{EFT}}

We rewrite the prediction for $D(Q)$ from the large-charge \ac{EFT}, \cref{eq:D-Q-ON-large-charge}, as
\begin{align}
  f(Q^{-1}) &\equiv Q^{-3/2}  D(Q) \\
            &= \cc32 + \cc12 Q^{-1} + c_0 Q^{-3/2}  + \Order(Q^{-2}).
\end{align}
The data for $f(Q^{-1})= Q^{-3/2}D(Q)$ as a function of $Q^{-1}$ should be linear close to $Q^{-1}=0$.
Therefore, close to the origin we can perform a linear fit to extract the leading coefficients $\cc32,$ and $\cc12$.
We can also include higher-order terms to get an estimate of systematic errors in the fit.
In \cref{fig:ON-large-charge-fits}, we show the data and fits
for all the $O(N)$ models considered in this paper ($N=2,4,6,8$).
We plot $Q^{-3/2}D(Q)$ vs. $Q^{-1}$ and perform fits of this data to the form
\newcommand\kmax{k_{\text{max}}}
\begin{align}
  f(x) = \cc32 + \cc12 x^{1} + c_0 x^{-3/2} + \sum_{k=2}^{\kmax} c_{\sfrac32 - k} x^{k}
\end{align}
where the power-series is cut off at $k=k_{\text{max}}$.

We note that the very small curvature of the data in \cref{fig:ON-large-charge-fits} indicates that the large-charge expansion is quite accurate even at low charge $Q$, and therefore low values of $\kmax$ are sufficient to obtain a very good fit.
In other words, $c_0$ and higher order coefficients are quite small.
Since it is not \emph{a priori} clear which values of $Q$ should be included in the fits, we perform various fits for the total charge in a window $Q \in [\Qmin, \Qmax ]$.
We fix $\Qmax=10$ and vary  $\Qmin$ over a small range.
As can be seen from the fits, even very small $Q$ seem to be described well by the ``large-charge'' expressions.
This is an intriguing phenomenon that was also observed by Refs.~\cite{banerjee_conformal_2018, banerjee_conformal_2019} for the $O(2)$ and $O(4)$ models.
As evidenced by \cref{fig:ON-large-charge-fits}, this observation continues to hold for larger $N$, which is interesting.

We also vary the truncation order $\kmax$ to get a sense of the systematics.
As might be expected, we find that excellent fits are obtained already with $\kmax \sim 2$  for most cases.

\begin{figure*}[htb]
  \centering
  \includegraphics[width=0.99\linewidth]{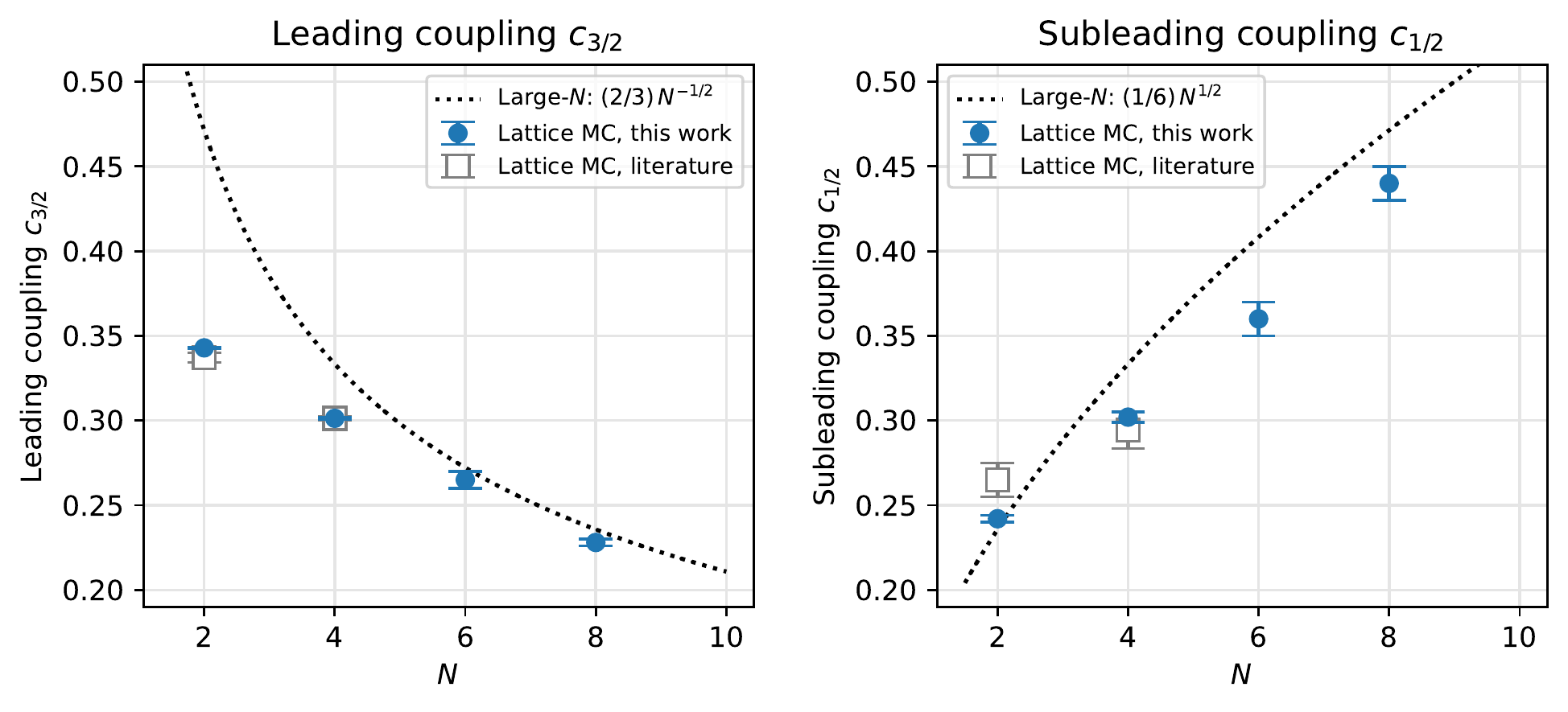}
  \caption[Large-charge EFT coefficients: lattice vs analytic large-$N$]{
    The two leading coefficients $\cc32$ and $\cc12$ in the large-charge expansion for the $O(N)$ Wilson-Fisher \ac{CFT}, shown as a function of $N$.
    The dotted lines show the large-$N$ predictions from Ref.~\cite{alvarez-gaume_large_2019}.
    The dark filled circles show the results of this work. 
    The unfilled squares show the values of the \acp{LEC} obtained in Refs.~\cite{banerjee_conformal_2018, banerjee_conformal_2019} for $N=2,4$.
    We find that the $c_{3/2}$ prediction is in excellent agreement with lattice \ac{MC} for $N \geq 6$.
    However, the subleading coefficient $c_{1/2}$, while within $\sim 3\sigma$ for $N\geq6$, seems to be in mild tension.
    Regardless, large-$N$ correctly predicts the qualitative trends for both couplings. 
  }
  \label{fig:large-NQ-results}
\end{figure*}

\subsection{Comparison with large-charge and large-$N$ results}

Finally, having extracted the \acp{LEC} of the large-charge \ac{EFT} ($\cc32, \cc12, \dotsc$), we can compare the results against the large-$N$ predictions \cite{alvarez-gaume_large_2019a} as well as earlier lattice calculations \cite{banerjee_conformal_2019, banerjee_conformal_2018}.
We show this comparison in \cref{fig:large-NQ-results} and \cref{tab:results}.
The solid lines are predictions from a large-$N$ prediction of Ref.~\cite{alvarez-gaume_large_2019}, while the results for $O(2)$ and $O(4)$ from previous lattice calculations \cite{banerjee_conformal_2019, banerjee_conformal_2018}.
We also show the comparison with the earlier literature on the extraction of these couplings.

First, we note that the qualitative prediction of large-$N$ that $\cc32$ decreases with $N$ while $\cc12$ increases is in excellent agreement with the lattice data.
Large-$N$ predicts that the two couplings are equal ($\cc32=\cc12$) at $N=4$. Interestingly, we find this to be true for the lattice results as well, although the actual numerical value of these couplings from large-$N$ differs from lattice by $\sim\si{10}{\%}$ at $N=4$.

On the quantitative side, we find that the leading coupling $\cc32$ seems to be in beautiful agreement with large-$N$ already by $N=6$.
On the other hand, the precise numbers for subleading coupling $\cc12$ suggest some mild tension.
Being a higher-order coupling, the error bars are larger as well.
The source for this discrepancy could be either that large-$N$ takes longer to converge for the subleading coupling, or simply that the lattice computations need to be performed at higher precision. At this stage, it is hard to draw any quantitative conclusions for $\cc12$.
It is interesting that the authors of Ref.~\cite{alvarez-gaume_large_2019} also noted a small puzzle regarding the subleading coupling $\cc12$, which might relate to this discrepancy as well.

We also compare with the values of the couplings obtained in the literature and find excellent agreement. For $N=2$, the couplings were obtained in Ref.~\cite{banerjee_conformal_2018} using the traditional lattice $O(2)$ model.
Their model was substantially different from ours, therefore our results provide an independent verification of their work.
As can be seen in \cref{fig:large-NQ-results}, both the leading and subleading couplings agree within $\sim2\sigma$.
For $N=4$, our technique is quite similar to the one used in Ref.~\cite{banerjee_conformal_2019}, where the authors also used a qubit $O(4)$ model.
While the agreement is expected, the qubit model used in Ref.~\cite{banerjee_conformal_2019} differs slightly from the one used here.
In particular, their model allowed a bond to `turn back on itself', whereas in our model there are no such `double bonds.'
Since we are working at the criticality, such details are not expected to matter.
This is indeed what we find.
In both these cases ($N=2,4$), our results agree with and improve over the previous results.

Finally, we remark that the next-higher-order coupling $c_0$ is predicted to have a precise value from the large-charge expansion. However, being higher order, it is very tricky to extract from the precision obtained in this work.  While it would be quite satisfying to extract the coupling $c_0$ numerically, in this work, we simply fixed it to the predicted value and let the other couplings vary.

\setlength{\tabcolsep}{3.5pt}
\sisetup{table-number-alignment = center, table-format = 2.6, table-align-text-post = false }
\ctable[
label = tab:results,
caption = Results for the large-charge expansion coefficients extracted from lattice Monte Carlo runs this work. We also include the numbers for the $O(2)$ and $O(4)$ models from literature. ,
mincapwidth=\textwidth,
center,
star,
]
{ l S S S S @{\extracolsep{1em}} S S S S S}{%
  \tnote[a]{Ref.~\cite{banerjee_conformal_2018}, traditional lattice $O(2)$ model}
  \tnote[b]{Ref.~\cite{banerjee_conformal_2019}, qubit $O(4)$ model}
  \tnote[c]{Ref.~\cite{alvarez-gaume_large_2019}, combined large-$N$ and large-charge expansion}}%
{\FL
  & \multicolumn{4}{c}{$\cc32$} & \multicolumn{4}{c}{$\cc12$} & \NN
  \cmidrule{2-5}
  \cmidrule{6-9}
  $N$ & 2 & 4 & 6 & 8 & 2 & 4 & 6 & 8 & \NN
  This work & 0.3429(5)& 0.3013(4)& 0.265(5)& 0.228(2)& 0.242(2)& 0.302(3)& 0.36(1)& 0.44(1) & \NN
  Large-N\tmark[c] & 0.471 & 0.333 & 0.272 & 0.236 & 0.236 & 0.333 & 0.408 & 0.471 & \NN
  Literature & 0.337(3)\tmark[a] & 0.301(1)\tmark[b] & {-} & {-} & 0.27(1)\tmark[a] & 0.29(1)\tmark[b] & {-} & {-}  &\LL
}

\section{Conclusions}
\label{sec:conclusions}

To explore the validity of the large-charge expansion for the $O(N)$ \ac{WF} \ac{CFT}, we performed lattice \ac{MC} computations for the $O(N)$ model at $N=2,4,6,8$.
In order to avoid a signal-to-noise ratio problem as we go to higher charges, we used a qubit $O(N)$ model, which was shown recently to have a second-order critical point in universality class of the $O(N)$ \ac{WF} fixed-point \cite{singh_qubit_2019a}.
This model allows for efficient lattice computations using a worldline formulation with a worm algorithm, and allows precise extraction of the conformal dimensions up to $Q=10$ and $N=8$.

Having computed the conformal dimensions, we then performed a fit to the prediction from the large-charge \ac{EFT} and extracted the two leading \acp{LEC} $\cc32$ and $\cc12$.
In line with what the authors of Refs.~\cite{banerjee_conformal_2018, banerjee_conformal_2019} observed for the $N=2,4$ models, we find that the large-charge expansion describes the data very well even for very small $Q$ for larger $N$ as well.
This is an intriguing fact about the large-charge expansion which would be nice to understand theoretically.

We finally compared our numerical results for the \acp{LEC} with a recent prediction from a combined large-charge and large-$N$ expansion \cite{alvarez-gaume_large_2019}.
We find that the large-$N$ prediction for the leading coefficient $\cc32$ agrees very well with the numerical computations already for $N \geq 6$.
The qualitative trends for both the leading \acp{LEC} $\cc32, \cc12$ are also predicted correctly by the large-$N$ expansion.
However, there seems to be a small tension between numerical values and large-$N$ for the subleading coefficient $\cc12$.
This can be either due to the fact that subleading coefficient is harder to extract numerically and there might be unresolved systematics, or it could be that we just need to go to larger $N$ for the subleading coefficient.
The authors of Ref.~\cite{alvarez-gaume_large_2019} also note a small puzzle regarding the $\cc12$ coupling, which might be related to this issue.
This merits further study and is left for a future publication.

\section*{Acknowledgments}

I would like to thank Shailesh Chandrasekharan, Susanne Reffert and Domenico Orlando for important discussions.
I would also like to like to thank Andrew Gasbarro, Hanqing Liu, Mendel Nguyen, Ronen Plesser, Roxanne Springer for useful conversations.  

The material presented here was funded 
in part by the DOE QuantISED program through the theory  consortium ``Intersections of QIS and Theoretical Particle Physics'' at Fermilab with Fermilab Subcontract No. 666484,
in part by Institute for Nuclear Theory with US Department of Energy Grant DE-FG02-00ER41132,
in part by U.S. Department of Energy, Office of Science, Office of Nuclear Physics, Inqubator for Quantum Simulation (IQuS) under Award Number DOE (NP) Award DE-SC0020970, 
in part by U.S. Department of Energy, Office of Science, Nuclear Physics program under award No.~DE-FG02-05ER41368, 
and in part by the J.\,Horst Meyer Endowment fellowship from Duke University.

\bibliographystyle{apsrev4-2}
\bibliography{refs}

\appendix
\section{Exact calculations on a small lattice}
\begin{figure}[htbp]
  \centering
  \includegraphics[width=0.9\linewidth]{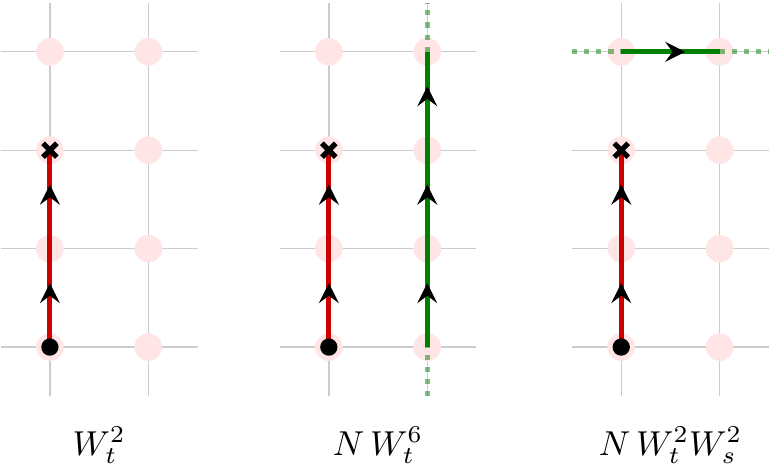}
  \caption{All the possible configurations with their weights on a $2d$ lattice of size $L=2$ and $\LT=4$ in the charge-$1$ sector. Here, we denote the weight of a temporal bond is $W_t$ and $W_s$ for clarity, although in our computations we have $W_t = W_s$. The dark circle represents a source and the cross represents a sink for the red worldlines. In our code the outgoing bonds at sources and incoming bonds at sinks are fixed to be in the temporal direction, so this completely fixes the red worldline on this small lattice. The other worldline can be of any color and orientation and contributes a factor of $N$ to the weights and second and third configurations. }
  \label{fig:exact-2x4}
\end{figure}

In this section, we test our \ac{MC} code by comparing against exact results for  a small lattice. We consider a $2d$ spacetime lattice with $L=2$ and $\LT=4$, where all possible  configurations can be easily enumerated.
We put exactly one source and sink of a fixed color.
In this case, there are exactly three types of configurations, as shown in \cref{fig:exact-2x4}.

Let the spacetime sites be indexed by integers $(x,t)$ where $x \in [0,1]$ and $t \in [0,3]$.  We place one source of a given color (say, red) at the site $(0,0)$ and a sink of the same color at $(0,2)$.
In \cref{fig:exact-2x4}, the sources are shown by a circles and the sinks are shown by a cross.
In our setup, the first outgoing bond from the sources is always frozen to be in the positive time direction, so that the outgoing bond always points up, and the incoming bond for the sinks is frozen to be coming in from negative time direction, so that the bond always comes in from below.  On this $2\times 4$ lattice, this completely freezes the worldline connecting the source and the sink.
Therefore, we can easily enumerate all the allowed configurations in this system, as shown in \cref{fig:exact-2x4}.
The first configuration shown has a single worldline of a fixed color and orientation since it connects a source with a sink, whereas the two configurations shown to its right have an additional worldline which can be of any of the $n$ colors and two orientations (so there are such $N=2n$ configurations).
The weight of each configuration is given by multiplying the weight of each temporal bond, $W_t$, and each spatial bond $W_s$. Therefore the partition function in the charge-$1$ sector is
\begin{align}
  Z_1 = W_t^2 + N W_t^6 + N W_t^2W_s^2.
\end{align}

We show results for the singlet density computed from the lattice \ac{MC} algorithm against exact results in \cref{tab:results-exact}.

\setlength{\tabcolsep}{7pt}
\sisetup{table-number-alignment = center, table-format = 1.2, table-align-text-post = false }
\ctable[%
label = exact_results, 
caption = {Comparison of the worm algorithm with exact calculations on a small $2\times 4$ lattice in the charge-$1$ sector. We show results for the ``singlet density'' observable, for various $N$, spatial weights ($W_s$) and temporal weights ($W_t$). We find complete agreement between the exact results and the worm algorithm. All the configurations contributing to this computation are shown in \cref{fig:exact-2x4}.} ,
label = tab:results-exact,
mincapwidth = \linewidth,
center,]%
{S S S S[table-format=1.7] S[table-format=1.4] S}%
{}{\FL
  $N$ & $W_s$ & $W_t$ & {Worm algorithm} & {Exact} & \ML
  1 & 0.62 & 0.47 & 0.49471(25) & 0.4947 & \NN
  2 & 1.75 & 0.47 & 0.38986(26) & 0.3899 & \NN
  3 & 3.00 & 0.11 & 0.379614(89) & 0.3795 &
  \LL
}

\end{document}